\documentclass[12pt]{article}
\usepackage{epsfig}

\def\Title#1{\begin{center} {\Large {\bf #1} } \end{center}}

\begin{document}

\Title{From confinement to adjoint zero-modes}

\bigskip\bigskip

\begin{raggedright}  

{\it Margarita Garc\'{\i}a P\'erez $^a$, Antonio Gonz\'alez-Arroyo $^{a,b}$ and
Alfonso Sastre $^{a,b}$  \index{GPGAS}\\
$^a$ Instituto de F\'isica Te\'orica UAM/CSIC, Modulo 8\\
$^b$ Departamento de F\'isica Te\'orica, Modulo 15\\
Universidad Aut\'onoma de Madrid\\
E-28049 Madrid, SPAIN}
\bigskip\bigskip
\end{raggedright}

\section{Introduction} 
Recently~\cite{adjoint_su} we have been embarked in a program aiming 
at  obtaining
the analytic expression and properties for the zero-modes of the Dirac equation in
the adjoint representation in the background of caloron
solutions~\cite{vanbaal,lee}.  Obviously, calorons are configurations that play
an important role in the semiclassical study of Yang-Mills theory at
finite temperature, and our zero-modes inherit this importance for the
case of Supersymmetric Yang-Mills theory. However, our work fits
into  a much broader context which has guided part of the work done by
our group in a period spanning more than a decade. It is precisely a
short promenade over this intellectual itinerary, as the title
reflects, what we are  presenting in this talk and contribution to the
conference. Although, a good deal  of the ideas reviewed can be found
in published papers, we find it useful to include them here, since the
ultimate goal has not been accomplished and some of them  are not
widely known.  

The starting point of our promenade is the problem of Confinement in
non-abelian gauge theories in four space-time dimensions. After
several general considerations about the problem itself, we review our 
proposal~\cite{frac} for a microscopic mechanism explaining confinement, 
which is based upon the presence of fractional topological charge instantons
in the Yang-Mills vacuum. Its merits and its problems are presented. 
In the following  section we review some of the results obtained in
the last decade which validate the term ``instanton quarks'' for these
fractional instantons. In particular, its connection to calorons is
explained. The next section, deals with adjoint zero-modes and how they 
fit into this program, to conclude reviewing some aspects of our
results for calorons. We end with some closing remarks. Bon Voyage! 
\section{Confinement}
Confinement is a fascinating non-perturbative property that we believe is shared by
QCD, the subject matter of this conference. Many  papers
have been written to give an explanation of this property and a proof
that it occurs in QCD. Very often, sterile conflict and debate
originates by semantic problems about what is meant by Confinement,
and what is there to be explained. Thus, we want to start this section
by discussing some aspects:

\vspace*{0.2cm}

\noindent{\bf What is Confinement?} 

Confinement in QCD is tied to the observation that there seems to be 
no asymptotic states having fractional baryon number, electric charge, 
and the rest of quantum numbers of quark fields. In seeking for an 
intuitive reason why this is so, one concludes that this must be due  
to the strength of the interaction among these fundamental fields. 
This idea led to the effective or macroscopic description of the 
phenomenon which was understood and explained in the 70's. 

\vspace*{0.2cm}

\noindent{\bf Effective theory of Confinement} 

The fundamental work of K. Wilson, G. `t Hooft, S. Mandelstam, and
others enabled to reach an understanding of the origin of the forces 
leading to Confinement. It was first established  that Confinement can
be defined in Yang-Mills theories (without quarks) as  a property
leading to a linear rise (with separation) of the potential between two
(non-dynamical) quark sources. Wilson's  lattice formulation of gauge
theories allows to analyze the problem from a classical statistical
mechanics point of view. The area law serves to characterize the
Confinement phase, which  turns out to be a typical phase of a gauge
theory. This contributed to lift what could be initially regarded as a
strange proposal by Weinberg, into a natural and presumably plausible possibility. 

In an abelian gauge theory, Gauss law and rotational invariance
predicts how the potential energy among two charges
depends upon distance. In two space-time dimensions linear confinement
takes place. In 3+1 dimensions we have the standard Coulomb law and
no confinement. However, if the electric flux is concentrated along
1-dimensional flux tubes instead, linear confinement takes place. No
matter how bizarre this possibility might look,  `t Hooft and Mandelstam 
realized that all one needs is a dual version of what happens with magnetic
flux in superconductors. 

\vspace*{0.2cm}

\noindent{\bf Microscopic mechanism of  Confinement} 

The previous point explains the nature  of the Confinement property.
It is the equivalent of the description of Superconductivity as the
result of the condensation of a charged field. However, this is not
the end. One still needs to explain why a particular system exhibits
this property and others don't. We dub this explanation, a {\it microscopic
mechanism for confinement}. For ordinary superconductors this is
provided by BCS theory. A similar explanation for high temperature
superconductors is still under debate. Equivalently, it is simple to
prove confinement in certain systems. For example, it is simple to 
compute the string tension in $Z_N$ gauge theories in a 2-d lattice,
in terms of the finite free energy of vortices. In 3-d vortices are
one dimensional and they are ineffective in leading to confinement at
zero temperature. However, in compact abelian gauge theories
Confinement in 3-d is obtained as a result of the finiteness of the
free energy of monopoles. In 4-d abelian gauge theories neither
vortices nor monopoles are effective in leading to Confinement at
sufficiently low temperatures. 

The previous considerations are firmly established and provide the
setting for discussing the occurrence of Confinement in 4-d (3+1)
non-abelian gauge theories. Many authors have proposed and attempted to
employ vortices and monopoles to prove Confinement for those theories. 
Any explanation should show how it works specifically for non-abelian
theories in the continuum limit. 

\vspace*{0.2cm}

\noindent{\bf Fractional topological charge mechanism for Confinement}
 
In a series of papers~\cite{frac}, our group proposed in the  90's a microscopic mechanism based 
on the role played by structures present in the Yang-Mills vacuum and
carrying fractional topological charge. These objects are specific of
non-abelian theories and are point-like in 4-d, as are vortices in 2-d and monopoles in 3-d.
Our proposal resembles strongly that made by Callan, Dashen and Gross
many year earlier~\cite{CDG}. These authors, however, based their proposal upon
the singular solutions, introduced by Alfaro, Fubini and Furlan~\cite{AFF}, called merons.
Our fractional charge  structures are non-singular.  

In our opinion, our proposal has certain appealing  features which we
will now list: 
\begin{itemize}
\item The existence of self-dual and smooth  classical configurations with lumps of
fractional topological charge $Q=1/N$ is firmly established
mathematically. These configurations have been analyzed  numerically in
great detail~\cite{frac_num}. They have infinite action but finite action density and
are local minima of the action. 
\item The counting of moduli  parameters describing self-dual  configurations
is provided by the index theorem. The result is that there are four
degrees of freedom for each $Q=1/N$ lump, which we associate with its
position. The space of self-dual configurations can be regarded as a
4-d gas of fractional instantons. 
\item One of the typical reasons why instantons are not considered
good candidates for confinement is because their free energy diverges
in the large N limit. Our configurations with fractional charge
escape this problem.
\item It is well-known that the QCD vacuum has a non-zero topological
susceptibility. In our model the same objects are responsible for this
phenomenon as for Confinement. This establishes a relation between the
topological susceptibility and the string tension which is in rough 
agreement with the data. 
\item This mechanism is  specific of 4-d non-abelian gauge theories.
\item Since classical Yang-Mills is a conformal theory fractional
instanton solutions exist in all sizes. However, quantum fluctuations
break the conformal invariance and end up determining the
characteristic size of these configurations in physical units. Like
for the case of instantons the small sizes are more improbable, while
large sizes are limited by the presence of neighboring structures. 
This ends up producing a fairly dense media, best described as a liquid.
\item It is known that  instantons do not produce Confinement. 
We gave both heuristic and numerical evidence that these configurations 
do indeed give a non-vanishing string tension. We prepared non-thermal
configurations containing an array of fractional instantons and
measured the Wilson loop in them. The linear potential term is seen
clearly in Fig.~\ref{fig:creutz}.

\item Our group was led to the idea of fractional instantons by the
study of the evolution of the Yang-Mills dynamics from small to large
periodic spatial volumes. The minimum energy in each electric flux sector 
was measured from small sizes, where semiclassical methods are precise,
to large volumes, where the infinite volume theory is recovered. For
twisted boundary conditions in space (non-zero `t Hooft magnetic flux), 
the evolution is smooth as shown in Fig.~\ref{fig:semiclass} (Left) and energies fuse into
the value dictated by the string tension. The Monte Carlo results match 
for small volumes with the computation from perturbation theory and
the semiclassical approximation. For the latter, the role of
fractional instantons is crucial. Fig.~\ref{fig:semiclass} (Right) shows a typical
configuration (after smoothening) displaying the presence of
fractional (and ordinary) instantons as a function of time.   

\item Early on the presence of fractional topological charge
configurations in the Yang-Mills vacuum was advocated by
Zhitnitsky~\cite{Zhitnitsky} to explain a puzzle in the  theta vacuum dependence with N.

\begin{figure}[htb]
\begin{center}
\centerline{
\epsfig{file=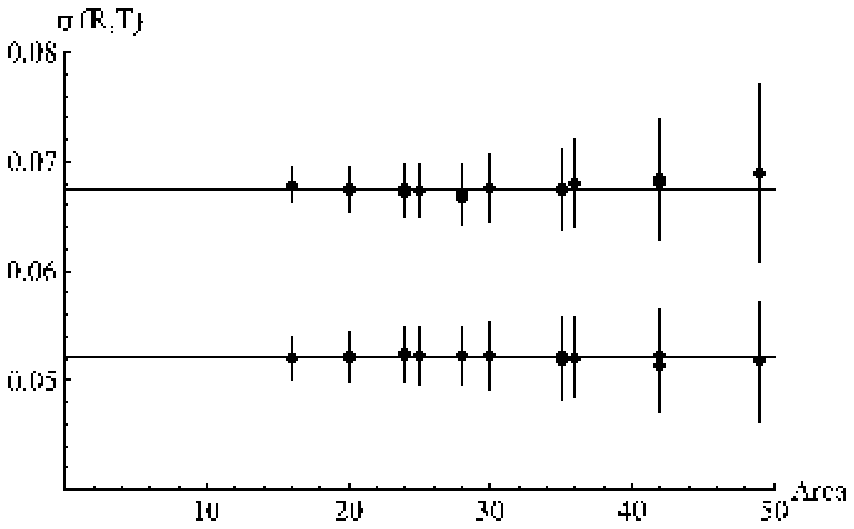,height=1.5in,width=2.5in,angle=0}\hspace*{0.4cm}
\epsfig{file=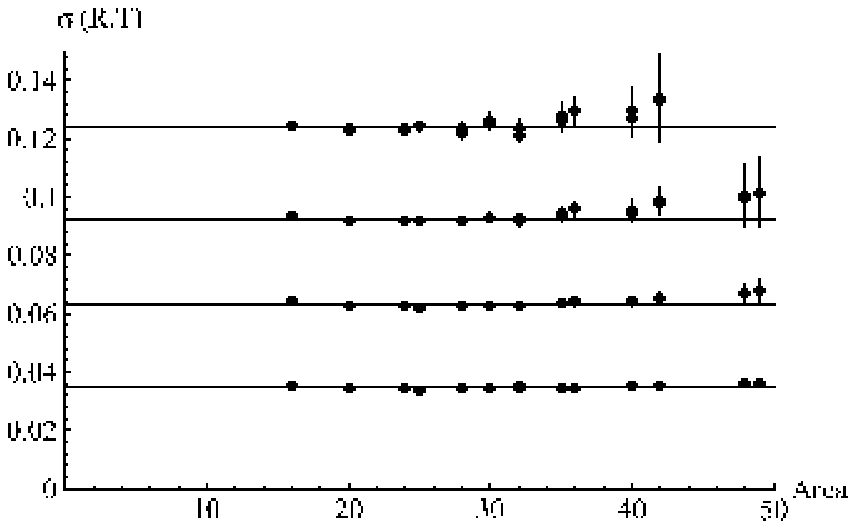,height=1.5in,width=2.5in,angle=0}}
\caption{The string tension vs the area of the Wilson loop measured
on non-thermal configurations containing an array of fractional instantons. 
Horizontal 
lines present results after different number of cooling steps (increasing from 
bottom to top).}
\label{fig:creutz}
\end{center}
\end{figure}
\begin{figure}[htb]
\begin{center}
\centerline{\hspace*{-0.3cm}
\epsfig{file=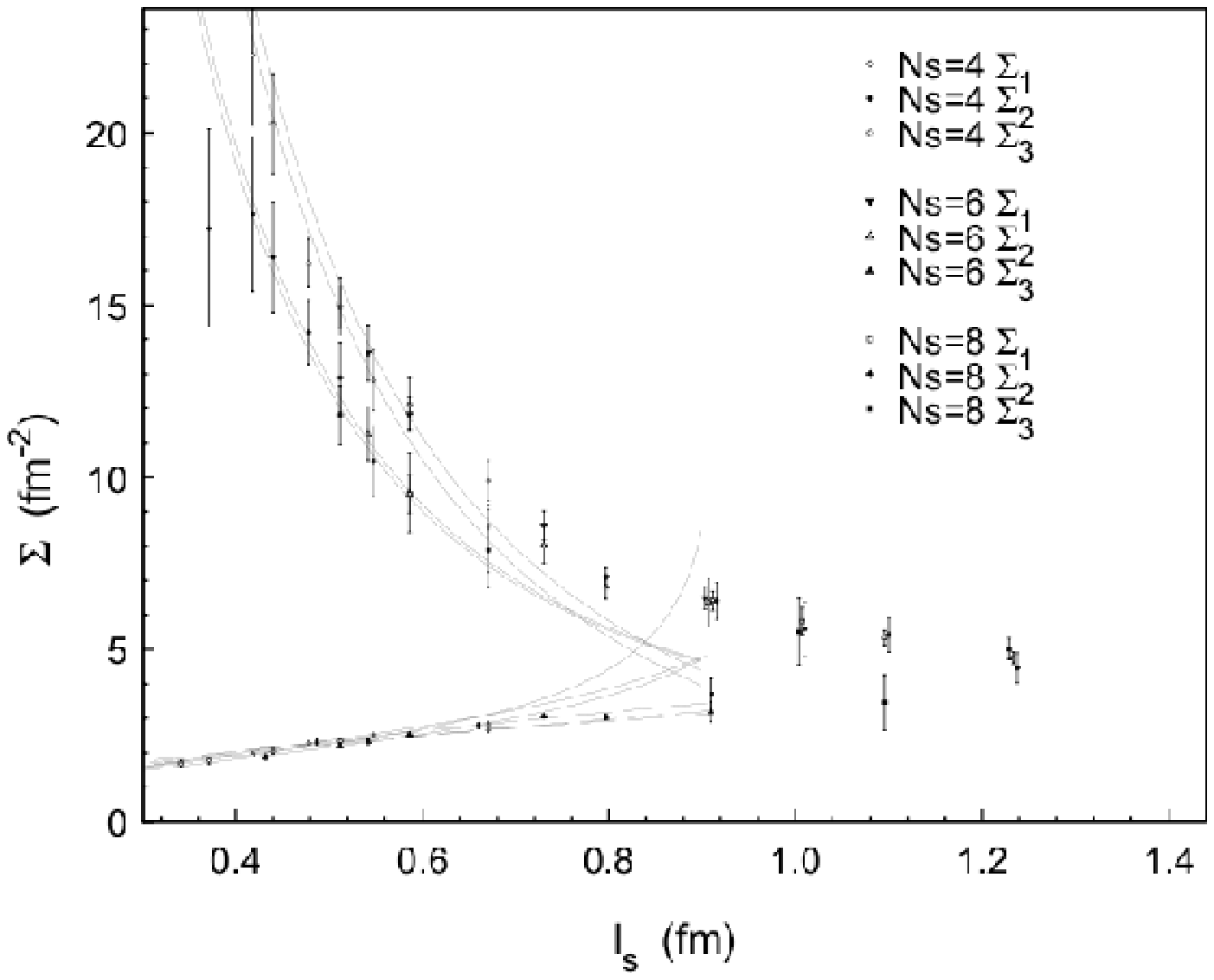,height=1.6in,width=2.6in,angle=0}\hspace*{0.4cm}
\epsfig{file=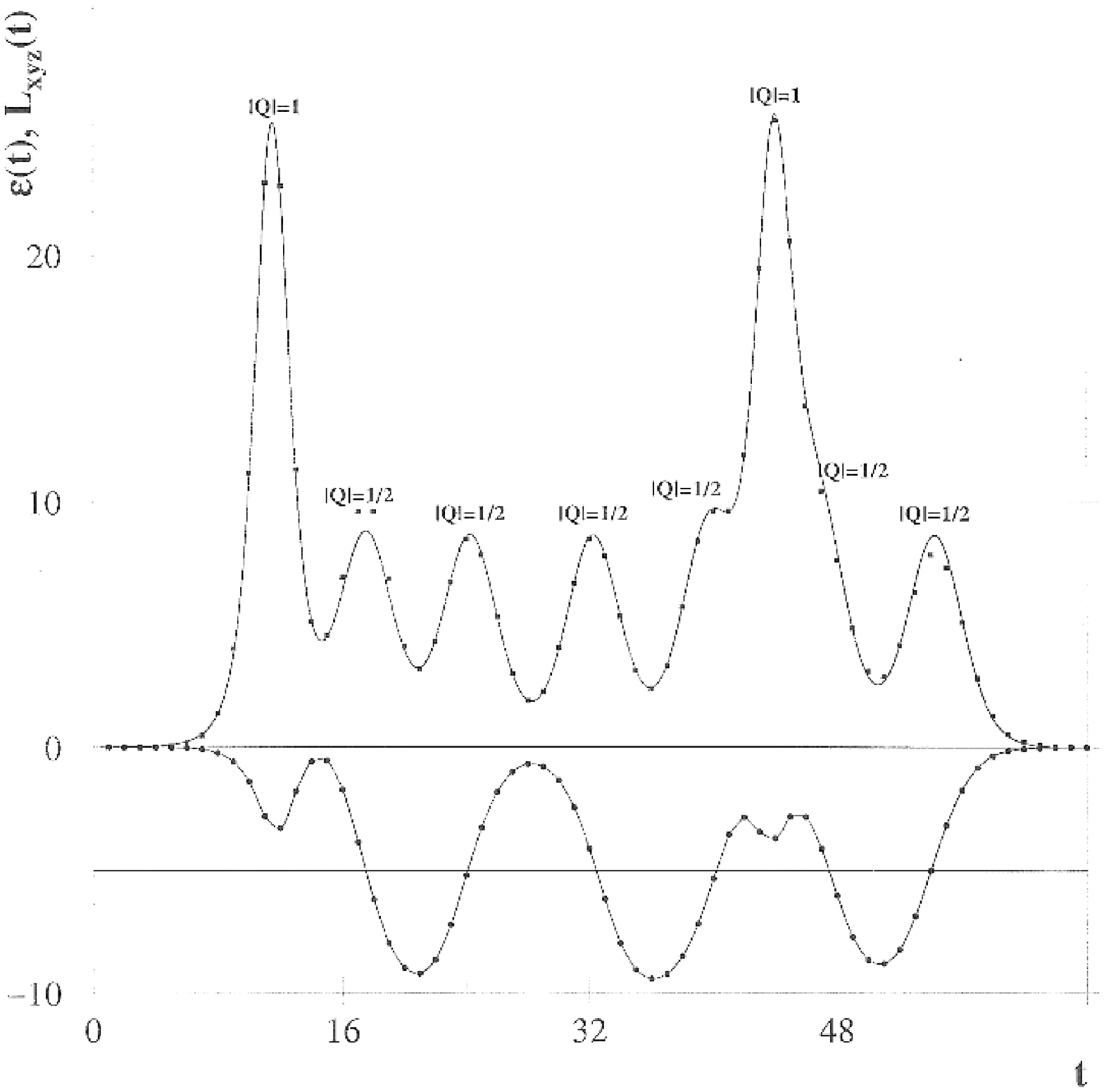,height=1.6in,width=2.6in,angle=0}}
\caption{Left: Effective string tension extracted from the electric flux 
energies as a function of the size of the twisted box. 
Right: Energy profile (top) and Polyakov line (bottom) of a typical
Monte-Carlo configuration for a small box.} 
\label{fig:semiclass}
\end{center}
\end{figure}
\end{itemize}

Despite all its appealing features, our program did not succeed in producing 
conclusive evidence of our description of the Yang-Mills vacuum 
and its corresponding explanation of the Confinement phenomenon
in non-abelian gauge theories in 4-d. The main reason is due to the
following difficulties: 
\begin{itemize}
\item Although we know that they exist and their shape and properties 
have been studied numerically, we lack an analytic formula for a
fractional instanton configuration. Calorons (see later) have ameliorated
the situation. 
\item As explained previously, the dynamical role of these
configurations, and the typical scales involved depend on its quantum
weights. Lacking a precise analytic formula for the configurations we
also lack an analytic handle over these weights. 
\item In any given  analytic description of the Yang-Mills vacuum one
would like to have an ansatz describing a  typical vacuum
configuration. In achieving this goal one faces enormous difficulties.
First of all, if the picture we describe is correct the dilute gas
approximation is not valid. A dense environment of fractional
instantons of equal sign could be described by complicated high topological
 charge
formulas. In addition, one has to face the problem that fractional
instantons of opposite signs should be present and the configuration
is not even a classical solution of the equations of motion. 

\begin{figure}[htb]
\begin{center}
\centerline{
\hspace*{-1.2cm}
\epsfig{file=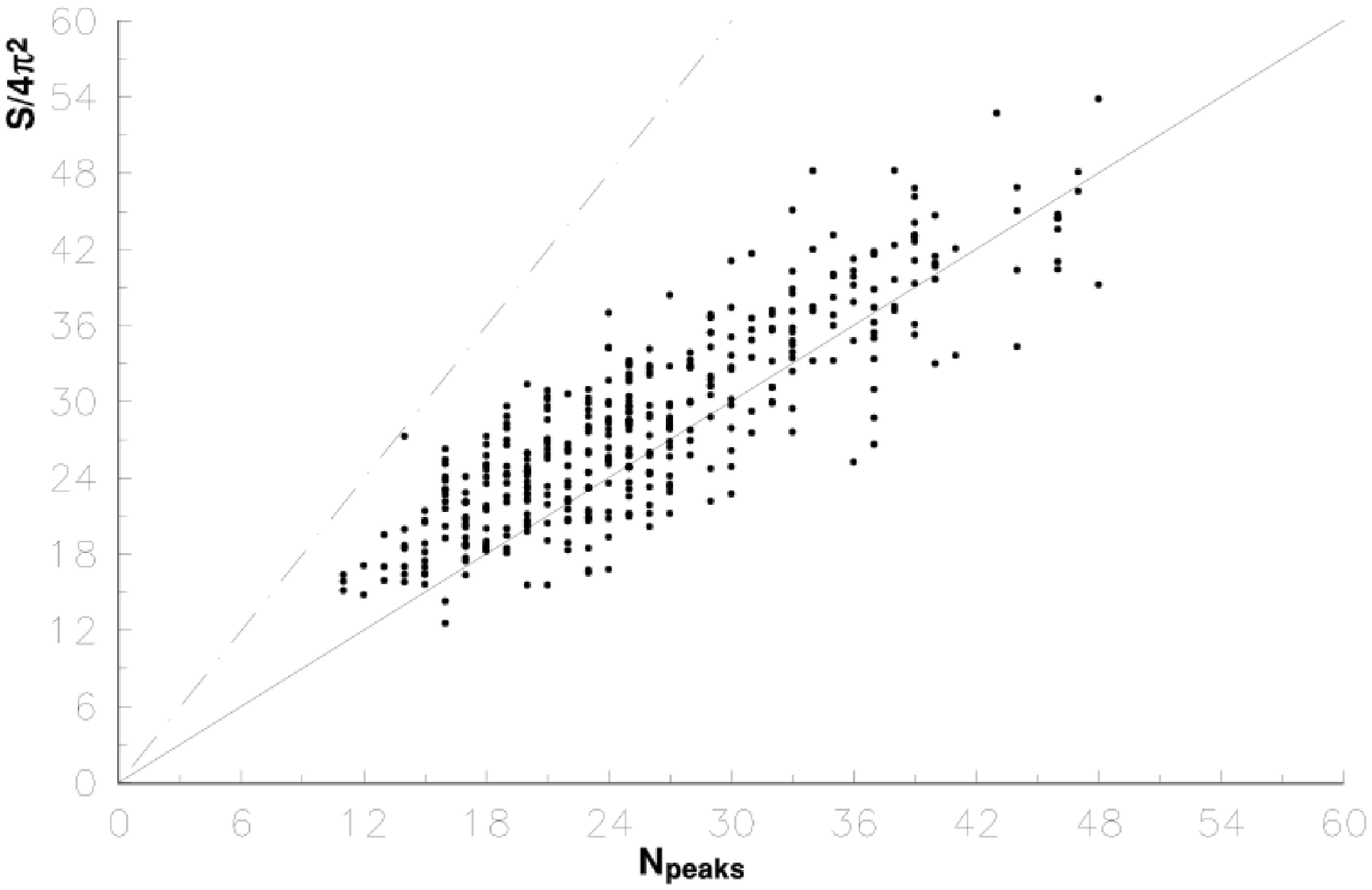,height=2.5in,width=2.5in,angle=270}\hspace*{1.2cm}
\epsfig{file=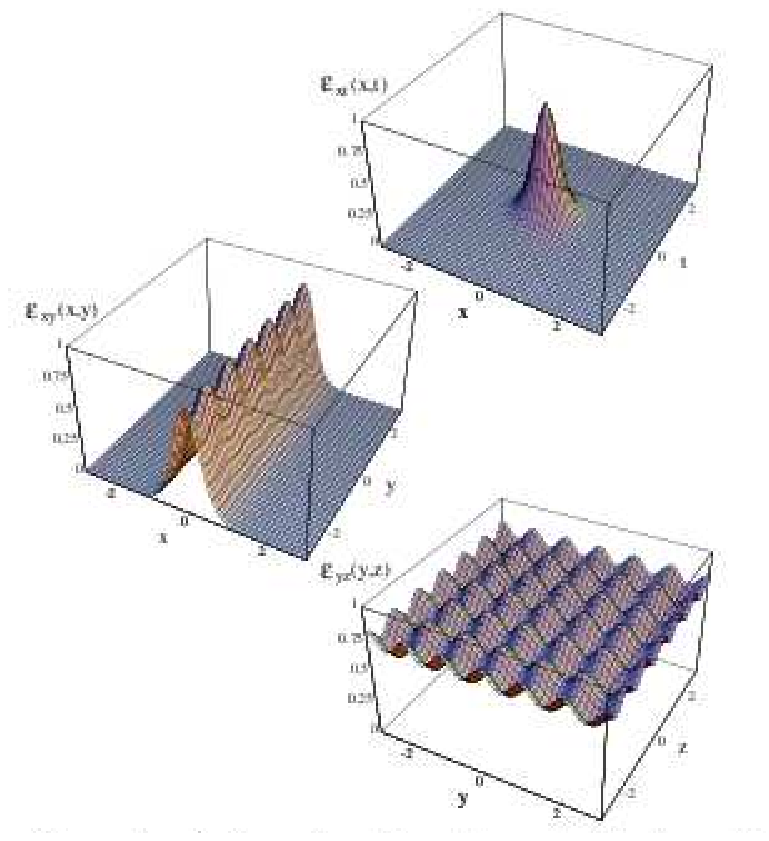,height=2.0in,width=2.0in,angle=270}}
\vspace*{-0.45cm}
\caption{Left: Scatter plot showing the action in units of $4 \pi^2$ vs the 
number of action-density peaks for several cooled Monte Carlo configurations.
Right: Several 2-d sections of the action density of the self-dual vortex 
solution.}
\label{fig:vortices}
\end{center}
\end{figure}

\item Ultimately, in order to check any proposal of this kind one
should try to resort to the analysis of Monte Carlo generated
lattice configurations. However, raw configurations tend to be very
rough as a result of ultraviolet divergences. A smoothening technique
is required. Although, we presented evidence that the lumps present in 
smoothened configurations carry fractional charge (See Fig.~\ref{fig:vortices} Left), 
this and other results have been questioned because they employed the
cooling method as a smoothening technique. 

\item Our work had also to face certain misunderstanding 
on the role played by twisted boundary conditions employed in some of
our papers. It should be realized that for large volumes 
the boundary conditions are inessential. However, this is not the case 
for small volumes. Twisted boundary conditions are more effective in showing 
fractional instantons in the low density regime. To understand the
situation one should compare with the similar situation taking place
in 1-d double-well scalar models. While kinks are crucial to understand
the dynamics at large volumes, antiperiodic boundary conditions are
more efficient than periodic ones in approaching this limit from small
volumes. 
\end{itemize}
\section{Fractional instantons and related objects}
Although no analytic expression has been obtained for the 4d
fractional instantons, there has been considerable progress 
on other classical structures, {\bf calorons},  which are intimately 
connected to them.  

A caloron is a self-dual Yang-Mills configuration in $S_1\times R^3$.
In other words, it is an instanton-like configuration which is
periodic in one direction. In a series of papers~\cite{vanbaal,lee}, analytic
expressions were obtained for $Q=1$ calorons. These formulas show in
glory detail how for a wide range of parameters of the solution, these
objects decompose into $N$ (for SU(N) gauge group) lumps carrying
non-integer charges adding up to 1. This is precisely the dissociation
that we advocate, and exhibit numerically,  to take place in 4d. The 
individual lumps which make a caloron were called {\em constituent
monopoles} in Ref.~\cite{vanbaal}. The way in which the topological charge is
split into all its constituents is dictated by the holonomy 
of the configuration  $P_\infty={\rm diag}\{e^{i 2 \pi
Z_i}\} $.

The approach followed by Kraan and van Baal to obtain the analytic formulas for 
calorons makes use of the {\bf Nahm transform}. This is a transformation 
which maps SU(N) self-dual gauge fields with topological charge $Q$ onto
SU(Q) self-dual fields with topological charge $N$. The transformed
field $\hat A_\mu(z)$ is obtained as follows:
\begin{equation}
\hat A_\mu(z) = i \int dx\, \psi^\dagger(x,z) \frac{\partial }{\partial
z_\mu} \psi(x,z)
\end{equation}
where $\psi$ is the solution of the modified Weyl equation in the fundamental
representation 
\begin{equation}
\bar{\sigma}_\mu ( D_\mu -2 \pi i z_\mu) \psi(x,z) = 0
\end{equation}
One important property is that the Nahm transform is an involution. 
Furthermore, periodic gauge fields map onto periodic gauge fields with
inverse periods.

In the case of  $Q=1$ calorons, the Nahm transform gives 
rise to 1-d periodic abelian gauge fields
$\hat{A}_\mu(z)$, which are self-dual except at discontinuities 
located at  the eigenvalues of the holonomy  $z=Z_i$. After 
solving for $\hat{A}_\mu(z)$,  applying a Nahm transformation, 
 one obtains the caloron formulas,
which depend upon $4N$ real parameters, in agreement with the index theorem.
As mentioned previously the action density is generically arranged
into $N$ independent 
monopole lumps, whose mass (topological charge) is proportional to 
$Z_{i+1}-Z_i$. (The  fermion zero-modes in the fundamental
representation have also been studied \cite{fundamental}).

\subsection*{Instanton-quarks}
What connection is there between the constituent monopoles and the
fractional instantons of our model of Confinement? It was found~\cite{cal_lat} 
that the constituent monopoles of minimal holonomy calorons (${\rm Tr}(P)=0$)  
 can be obtained as 1-d arrays of our fractional instantons. This explains the
 periodicity  in 1-d. In this fashion the latter appear as the building blocks
 from which calorons are built. This substantiates the idea of instanton
quarks and identifies fractional instantons with them. Although,
instanton quarks cannot exist isolated they can be arranged into
self-dual (and hence classically stable) smooth structures. One-dimensional
arrangements can be identified with the world line of monopoles, which
as mentioned previously coincide with the constituent monopoles of
calorons. Similarly two-dimensional arrangements describe the world
sheet of non-abelian self-dual vortices which have been shown to exist 
(See Ref.~\cite{vortices} and Fig.~\ref{fig:vortices} Right). 

At zero-temperature we do not see any reason why there should be a
hierarchy among distances to neighboring fractional instantons, which 
should rather resemble a 4-d liquid.  Ultimately, the question of whether
instantons dissociate into instanton quarks and 
their possible tendency to arrange into 1,2,3, or 4-dimensional
structures is determined by the free energy (or quantum weight) of each type of
configuration. On general grounds entropy favors dissociation.
Recent results on calorons point in this direction as well~\cite{diakonov}. 

\section{Adjoint quarks as probes}

Recently the interest in the study of Yang-Mills theory with quarks in the
adjoint representation has boosted. Traditionally, most of the results were
focused in the case of gluino fields in Supersymmetric Yang-Mills theory. 
Lately other QCD-like theories have received a lot of attention, some of which
possess quarks in the adjoint. The reasons are varied, ranging from studies in
Technicolor,  large N reduction, and in general as testing
grounds  of several concepts of Yang-Mills dynamics.

There is another goal of using adjoint quarks,  which we have been concerned
with, and fits into the general program of elucidating the structure of the
Yang-Mills vacuum and the origin of Confinement. As mentioned previously, 
any proposal should be contrasted and checked versus configurations obtained
by Monte Carlo simulations on the lattice. The main difficulty  is how to 
extract the information about the structures contained 
in these configurations filtering the higher momentum fluctuations. 
Traditionally cooling and smearing algorithms have been used to derive smooth
enough configurations which can be analyzed for structures. Although the
methodology is generally accepted in extracting global quantities, as the
total topological charge, their usage to probe local structures has been 
criticized, since both mechanisms might modify the initial gauge configuration
in an uncontrolled way. 

In the last years, a new approach based on the Dirac 
operator has been employed~\cite{latindex}. The low-lying eigenmodes of the Dirac
operator can be regarded as observables of a pure gauge theory which are less 
sensitive to the high momentum fluctuations of the field. Furthermore, they
should serve to track the structures present in a given configuration 
and their topological nature. From this general idea to a particular proposal
one has to determine which eigenmodes will be used and how. It is also 
possible to apply this idea for quarks in any  representation, although most
of the work has focused upon the fundamental representation.
One difficulty seems to be that even if the configuration is smooth the
fermionic densities of the zero-modes do not exactly reproduce the shape of
the gauge action density. In particular, for instantons for example,  they
fall off with a different power. There is a notable exception to this 
for the case of the adjoint representation. This is precisely the idea that
inspired a new  proposal~\cite{adj_probes}, which is explained in the next
paragraph.
 
For any gauge field configuration,   $A_\mu$, which is a solution of the
classical equations of motion, and any constant 4-spinor $v$, there exists a 
 zero-mode of the Dirac operator in the adjoint representation
$\psi_{ss}^a = \frac{1}{8}F_{\mu\nu}^a\left[\gamma_\mu,\gamma_\nu\right]v$, 
whose density is proportional  to the gauge action density. This is the 
{\it supersymmetric zero-mode}. Furthermore, each of its chiral components 
reproduces the self-dual or anti-self-dual part of the action density. 
This  is not the only property that makes the supersymmetric zero-mode 
special.  A look at the previous expression shows that, for an appropriate
choice of $v$, ${\rm Im} (\psi_{ss}^a)_ 1= 0$ at all points of space. In
particular, for the Weyl representation of the Dirac matrices and positive
chirality, $v=(1,0,0,0)$ does the job.

\begin{figure}[htb]
\begin{center}
\centerline{
\epsfig{file=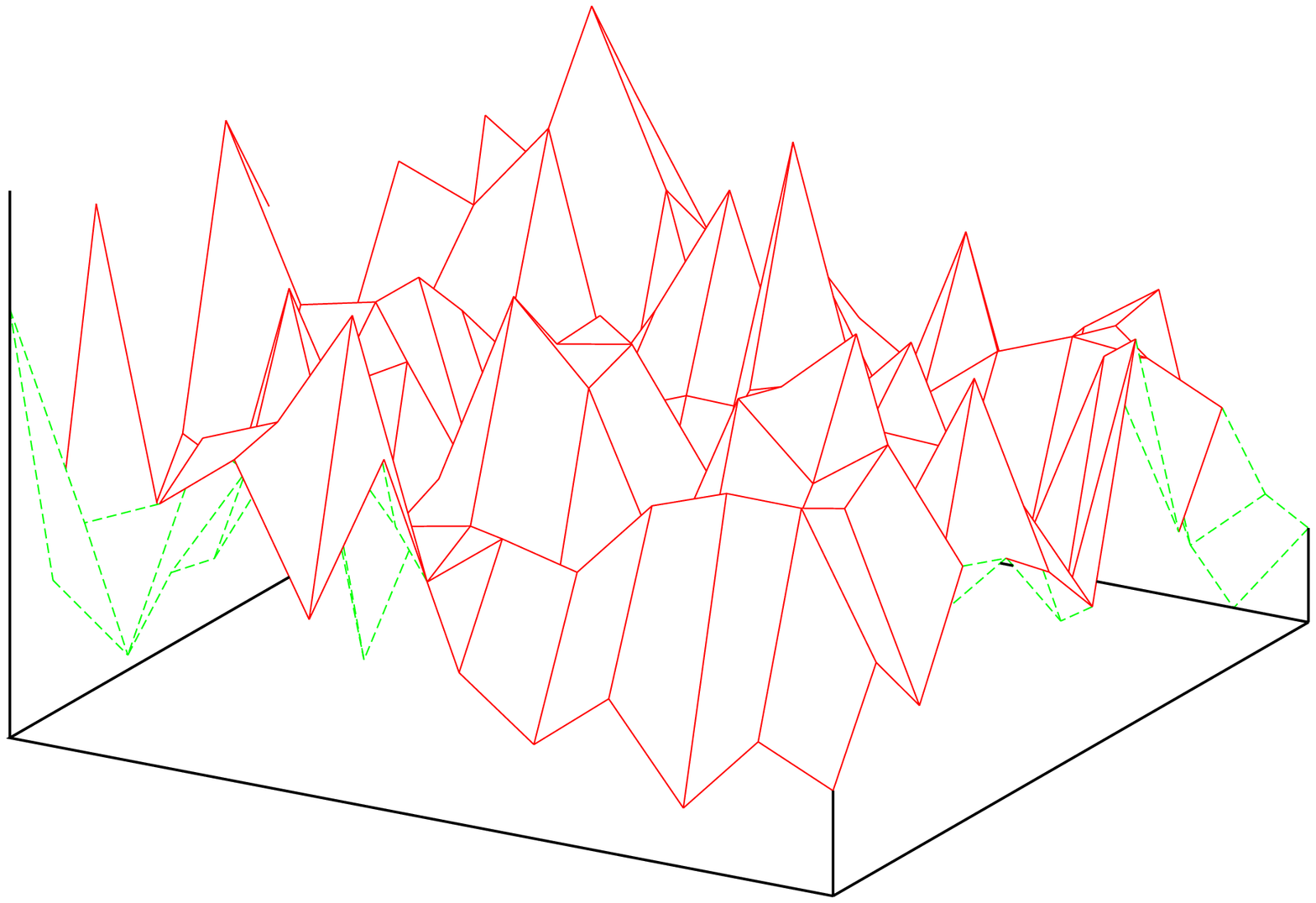,height=1.8in,angle=270}
\epsfig{file=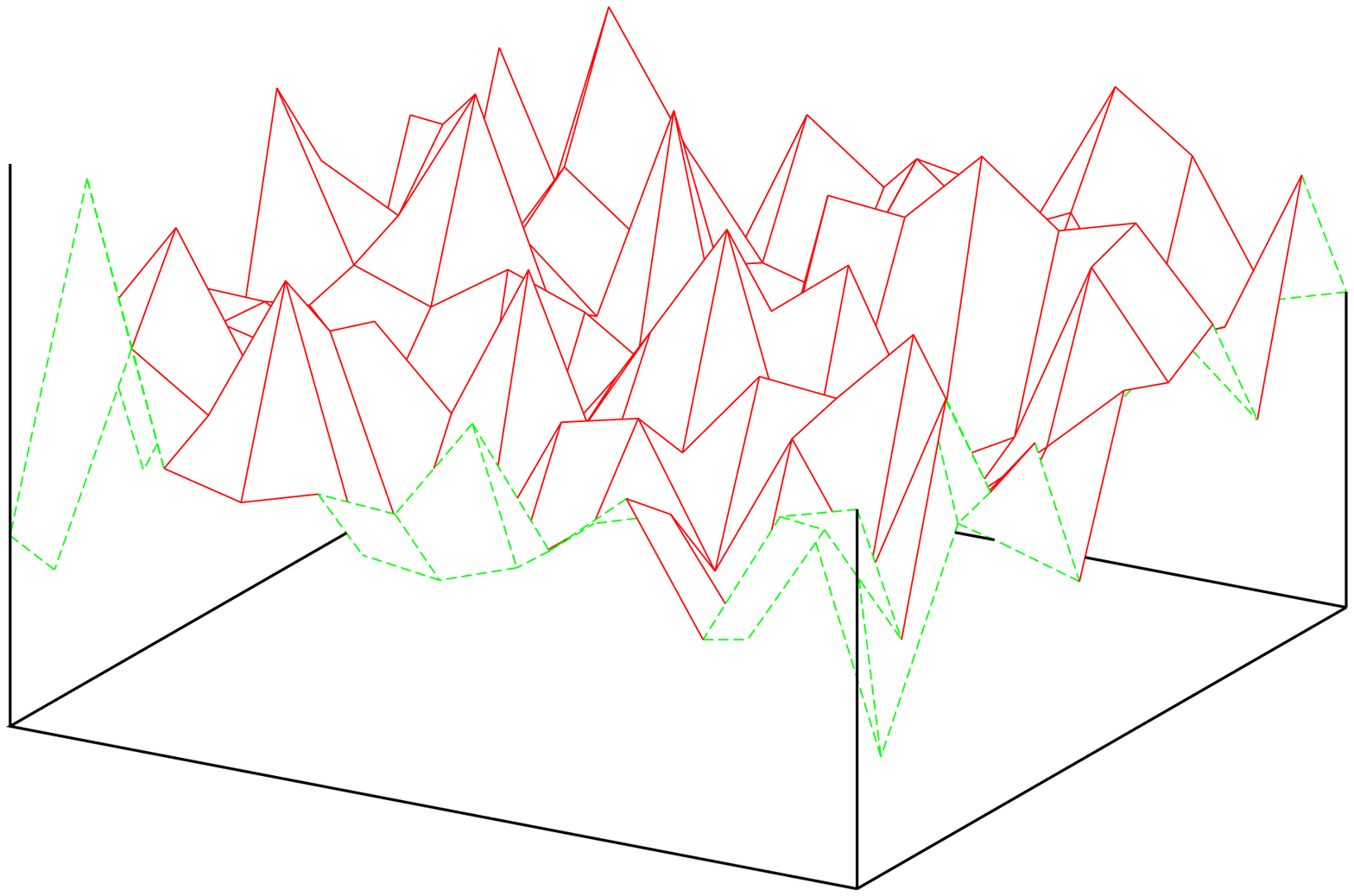,height=1.8in,angle=270}
\epsfig{file=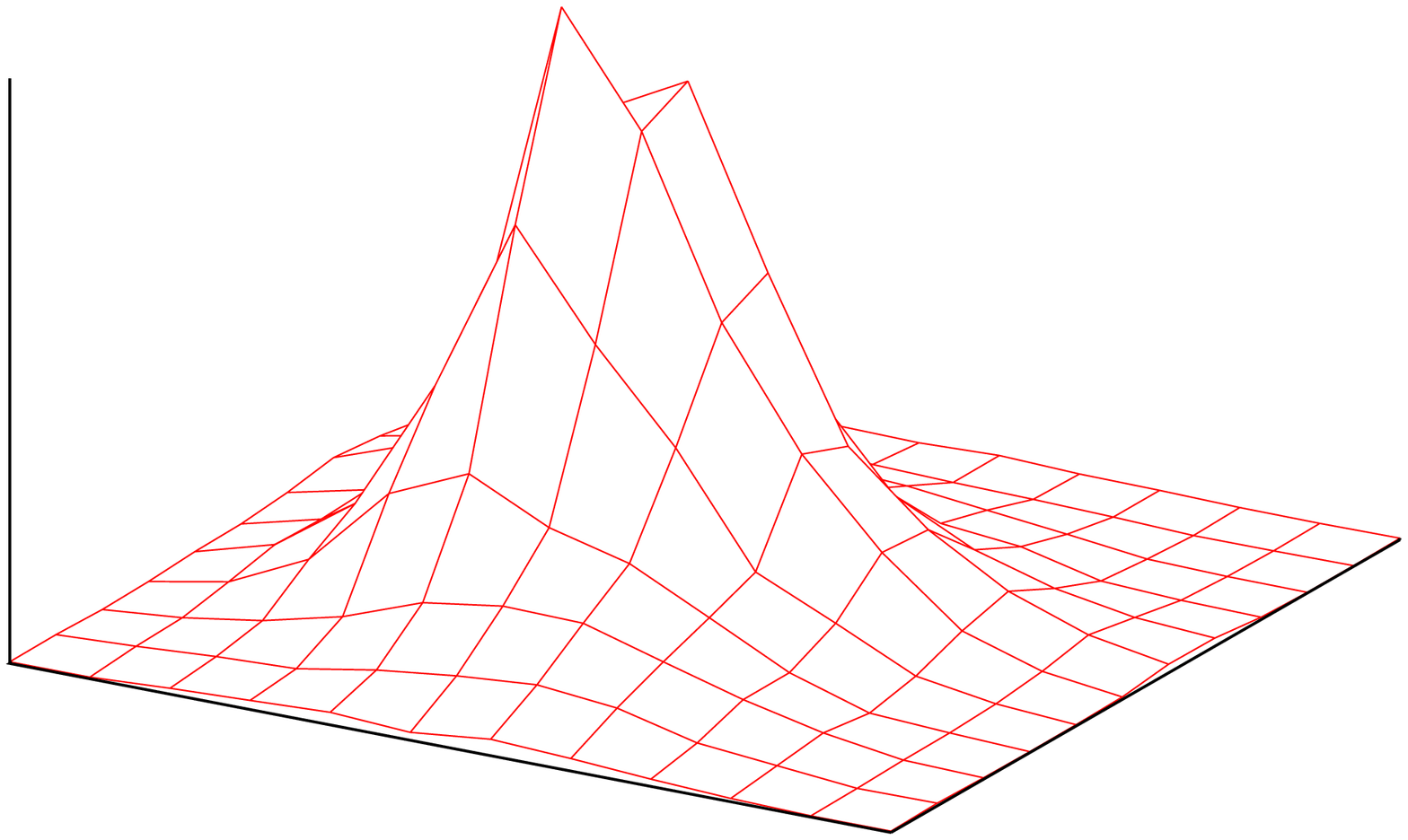,height=1.8in,angle=270}
}
\caption{From left to right: 2-d slice of the action, topological charge and susy zero-mode densities after  applying eight, $\beta\!=\!6.4$, heat-bath sweeps to 
an instanton.}
\label{fig:probe}
\end{center}
\end{figure}

The previous facts form the basis of our filtering method~\cite{adj_probes}. 
One selects the  quasi-zero mode of the adjoint Dirac operator in each chirality 
that satisfies the reality condition explained in the previous paragraph. 
The corresponding density provides a filtered version of the action density
and topological charge density. Practical implementations applied to
configurations on the lattice give promising results. As an example, suppose
that one starts from a smooth SU(3) Yang-Mills $Q=1$ configuration, an
instanton. If we now apply even  a small number of heat-bath sweeps at
$\beta=6.4$, the configurations becomes very  rough and neither the
action density nor the topological charge density reveal the underlying
instanton structure. This is displayed in the left and middle graphs of
Fig.~\ref{fig:probe}. On the right, the filtered action density is displayed, 
showing the instanton with the right size and location. The latter is obtained
as the density of the ground state of the operator $O_+$ obtained by
projecting the square of Neuberger's Dirac operator onto the space of positive
chirality vectors satisfying the reality condition.

\section{Adjoint zero-modes for calorons}

Motivated by the ideas explained in previous sections, 
our group has recently derived analytic formulae for the adjoint (gluino) zero-modes
in the background of $Q=1$, SU(N) calorons with non-trivial holonomy~\cite{adjoint_su}.
We have studied both periodic and antiperiodic solutions relevant for supersymmetric and 
finite temperature compactifications respectively.
Here we will present a brief summary of our results and refer the interested reader 
to the original papers for further details. 

We note that Weyl adjoint-zero modes appear in CP-pairs which can be arranged into a 
$2\times 2 $ quaternionic matrix: $\Psi= (\psi,\psi_C)$.
Our construction relies on the relation between solutions of the 
left-handed Weyl equation and self-dual deformations of the gauge 
field $\delta A_\mu$. It can be shown that
$\Psi\equiv\delta A_\mu \sigma_\mu$ is a solution of the adjoint Weyl equation, 
$\bar \sigma_\mu D_\mu \Psi=0$, provided that $\delta A_\mu$ satisfies the background
gauge condition ($D_\mu \delta A_\mu=0$).
This connection allows to obtain adjoint zero-modes by differentiating the gauge 
potential with respect to the parameters of the moduli space of solutions. 
As an example, the supersymmetric zero-mode discussed in the previous section can be
derived from the deformations of the gauge field associated to translations,
i.e. $\delta A_\mu = \partial_\rho A_\mu + D_\mu(-A_\rho) =F_{\rho \mu}$.
The construction provides zero-modes that are periodic in the thermal cycle with the 
same period as the gauge field $A_\mu$. In order to obtain antiperiodic solutions, relevant
for finite temperature, one has to resort to what we have called the {\it replica} trick.
It is based on the observation that antiperiodic solutions become periodic in the
double period and can thus be extracted from deformations of the $Q=2$ solution.
This {\it replica} procedure can be easily generalized to higher number of replicas 
and allows to obtain zero-modes with arbitrary periodicity.

Although the general caloron solution for arbitrary charge $Q$ and its moduli,
is not known, the self-dual deformations of the replicated caloron 
gauge field can be  derived  by making use of the Nahm-ADHM  formalism introduced in section 3. 
The Nahm-dual gauge field associated to the $L$-times replicated caloron 
is a one-dimensional U(L) gauge field given by $\hat A_\mu^R (z)= \rm{diag} 
(\hat A_\mu (z+(n-1)/L))$, $n=1,\cdots, L$, with $\hat A_\mu (z)$ the Nahm-dual of the 
$Q\!=\!1$ caloron gauge field. Without entering into details, it can be shown that
the problem of finding the adjoint zero-modes of the Dirac equation can be mapped
into the simpler one of solving the Nahm-dual adjoint Dirac equation, with gauge field 
$\hat A_\mu^R (z)$ and delta function sources at $z=Z_a+(n-1)/L$. 
The general solution of the latter for arbitrary number of replicas and 
the formulas linking them to adjoint zero-modes of arbitrary periodicity 
can be found in~\cite{adjoint_su} and will not be detailed here. 
Instead we will focus on discussing how the spatial structure of the zero-modes
relates to the location of the constituent monopoles. We will show how the results
fit nicely into the predictions of the index theorem~\cite{callias,unsal_index}.

\begin{figure}[htb]
\begin{center}
\centerline{
\psfig{file=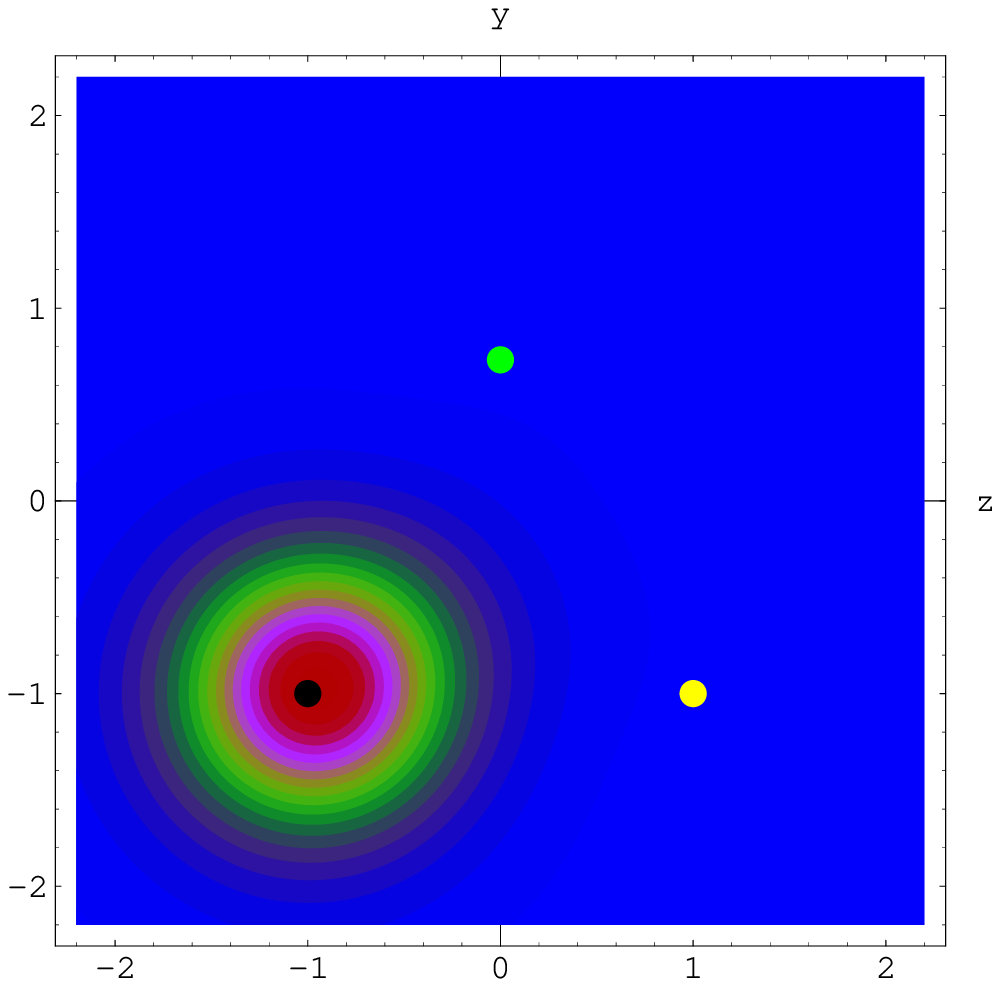,angle=0,width=3.5cm}
\psfig{file=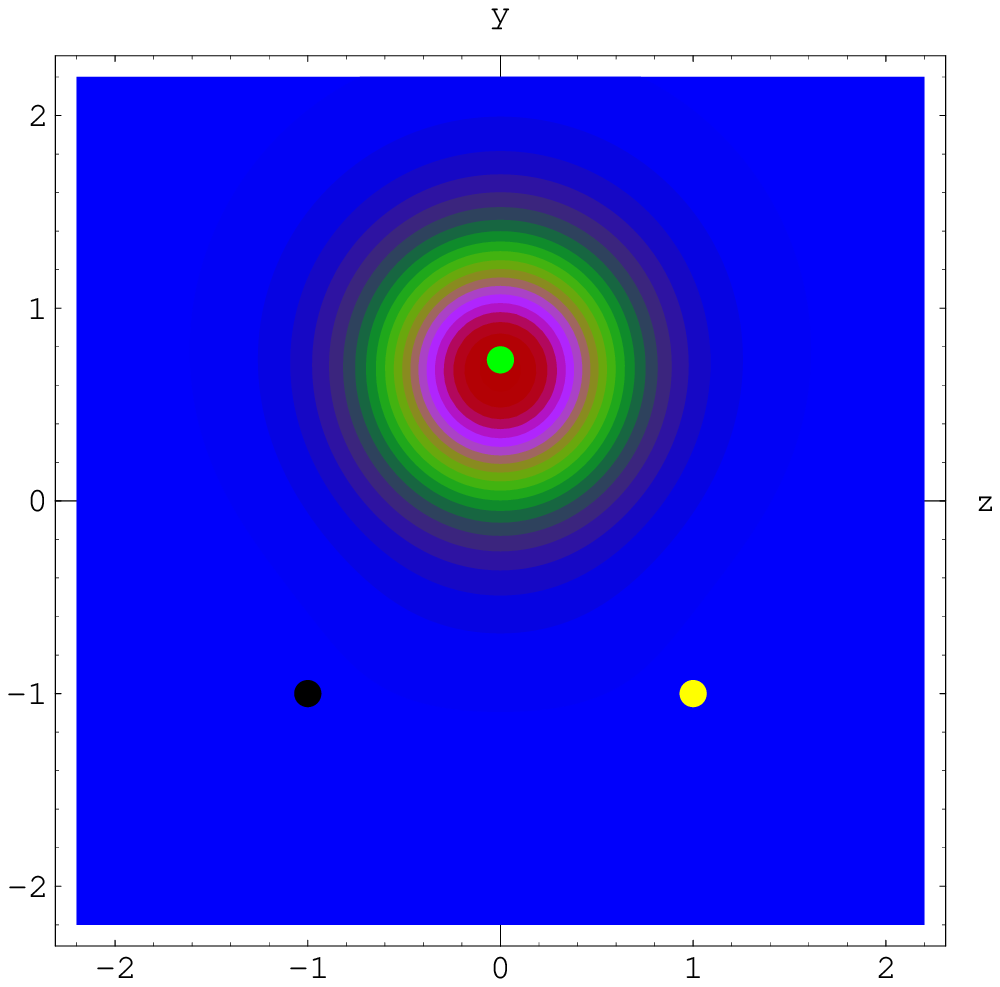,angle=0,width=3.5cm}
\psfig{file=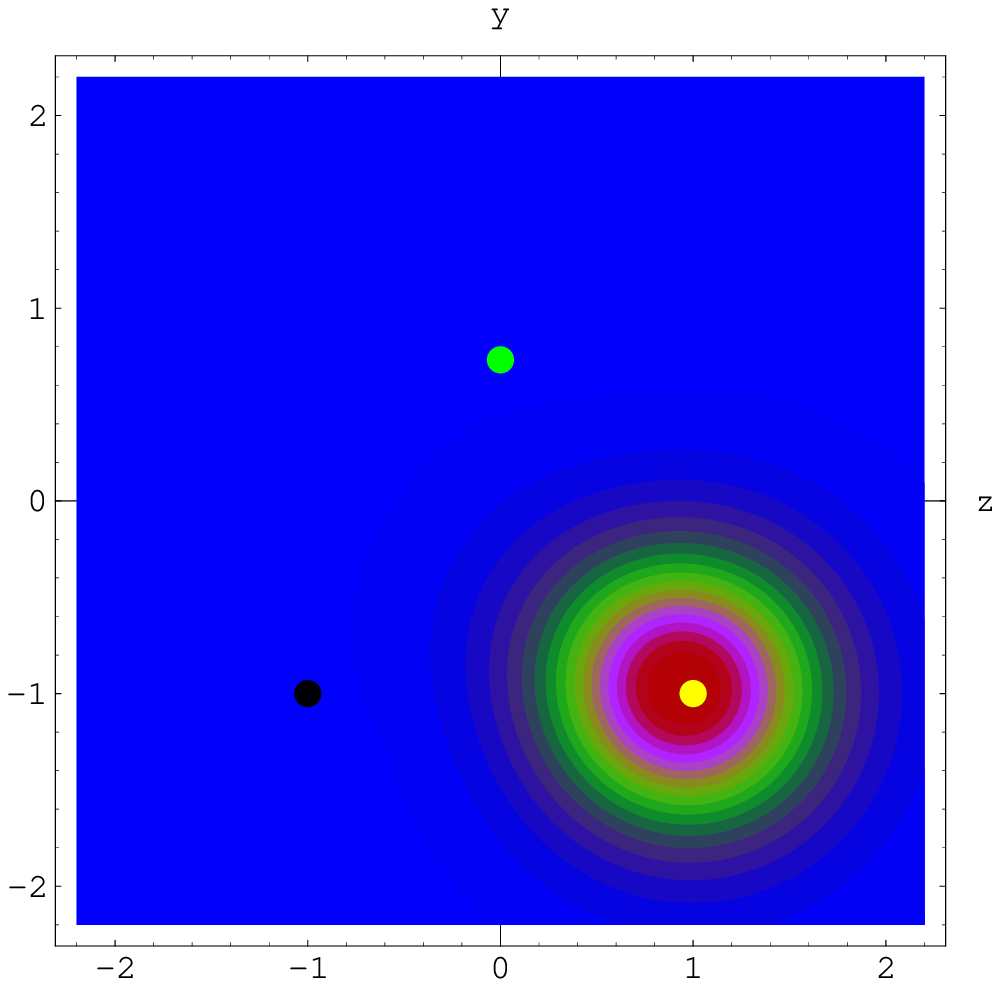,angle=0,width=3.5cm}}
\centerline{
\psfig{file=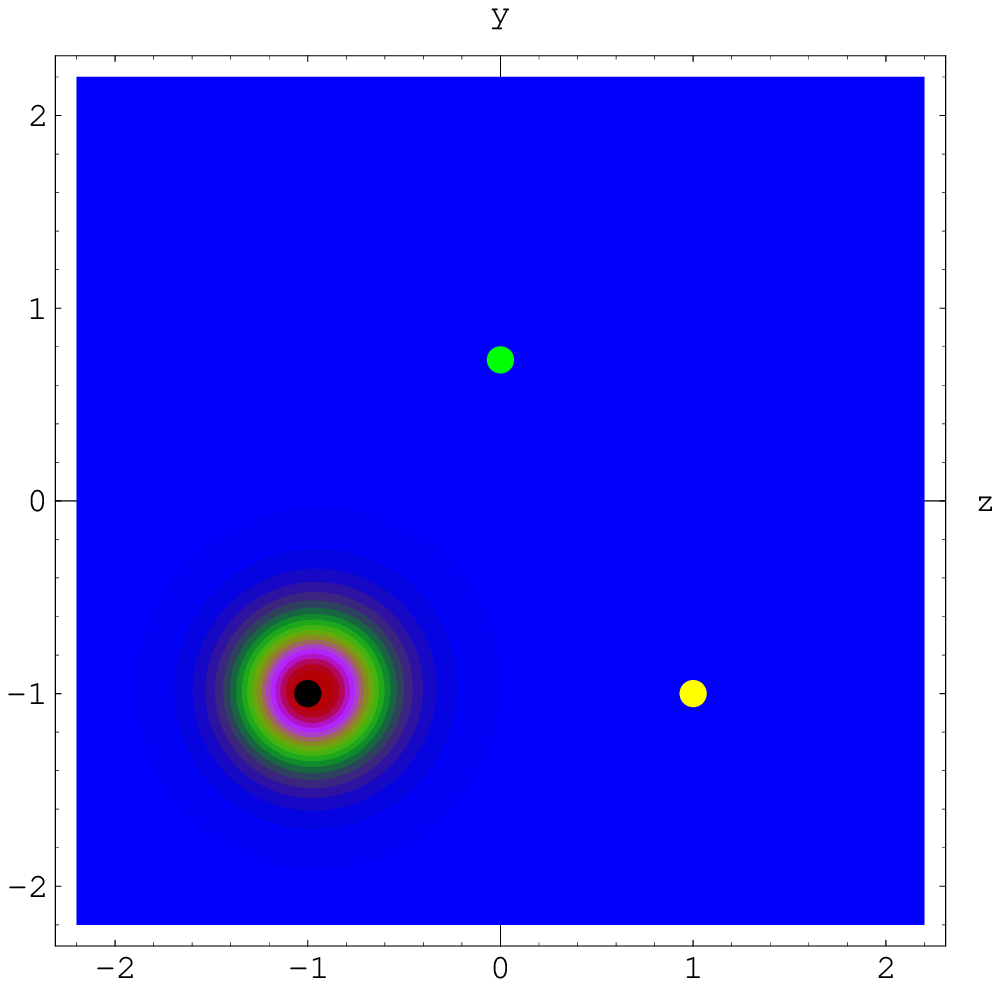,angle=0,width=3.5cm}
\psfig{file=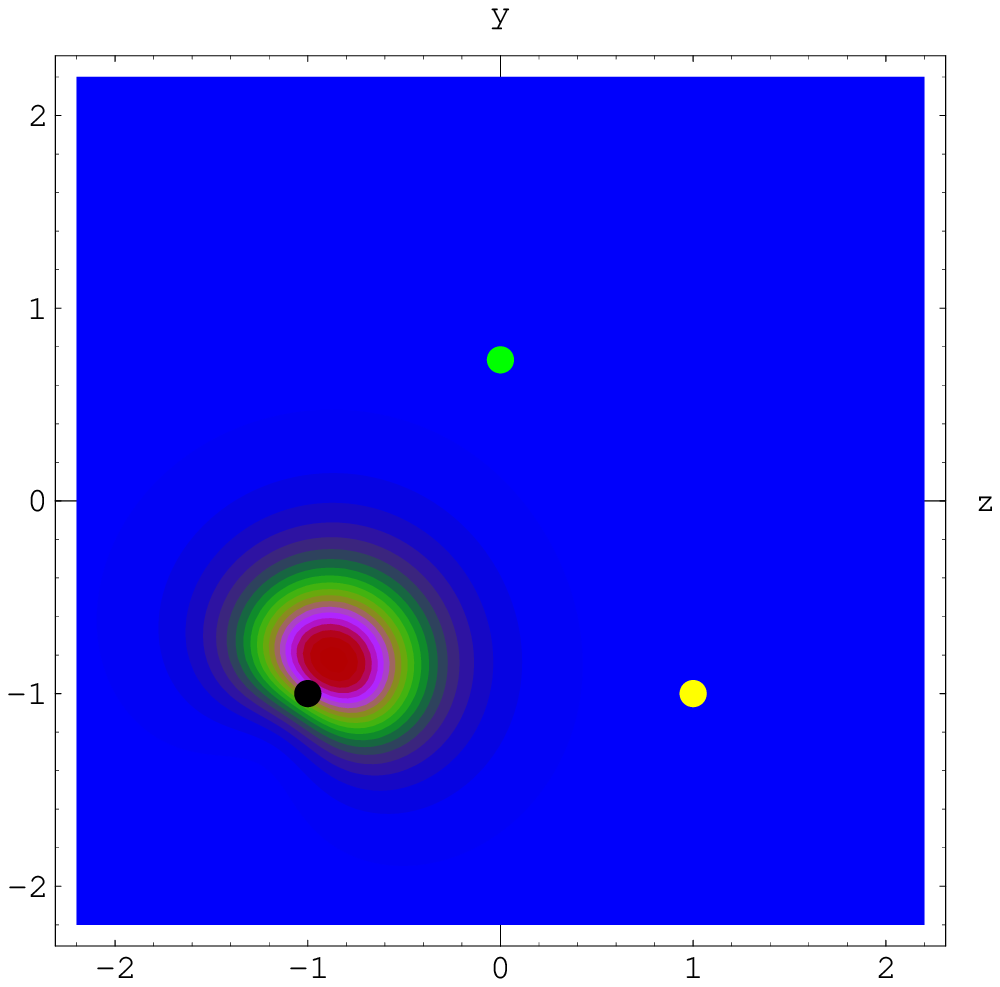,angle=0,width=3.5cm}
\psfig{file=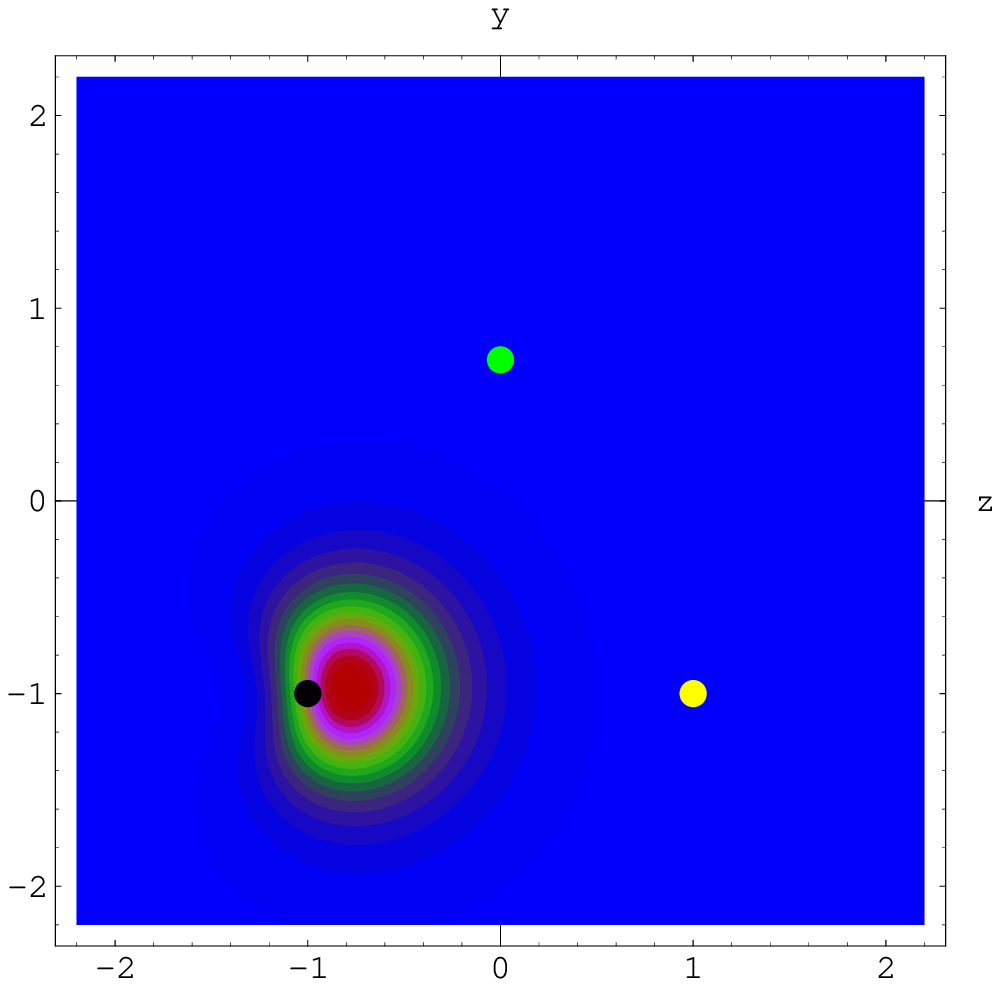,angle=0,width=3.5cm}}
\caption{Contour plots of the SU(3) antiperiodic zero-mode densities
at $x\!=\!t\!=\!0$.
The top (bottom) row corresponds to monopole masses:
$m_1\!=\!m_2\!=\!m_3\!=\!2\pi/3$ ($m_1\!=\!4\pi/3$, $m_2\!=\!m_3\!=\!\pi/3$).
The small filled circles indicate the monopole positions.}
\label{fig:su3_aper}
\end{center}
\end{figure}

In the limit of well separated constituent monopoles the structure of periodic 
zero-modes of the caloron is rather simple. There is a CP-pair of zero-modes
centered at each monopole, in accordance with the prediction of the Callias
index theorem~\cite{callias}. On the contrary, anti-periodic zero-modes
exhibit a more complicated pattern which we will illustrate for gauge group SU(3).
The number of adjoint zero-modes predicted by the index theorem for the SU(3) L-times 
replicated caloron is given by~\cite{unsal_index}:
\begin{equation}
I_{\rm adj} (L,n_1,n_2,n_3) = \sum_{a=1}^3 2 n_a  ( 3 q_a + \epsilon )\,,
\label{eq:adj_index}
\end{equation}
where the index $a$ parameterizes each of the constituent monopoles,
$\epsilon \equiv L-\sum_a q_a$, and 
$q_a\equiv [L m_a/2 \pi]$, with  $[v]  \equiv {\rm max}\{ n\in {\cal Z} \, | \, n\le v\}$,
and $m_a$ the masses of the constituent monopoles.   
For the L-times replicated caloron $n_a=1, \, \forall a$, and the total index 
correctly gives $6L$ adjoint zero-modes. According to the formula, out of those there
are $2 ( 3 q_a + \epsilon )$ associated to the $a$-th constituent monopole.
Periodic zero-modes correspond to $L=1$ for which $I_{\rm adj} = \sum_{a} 2 n_a$, 
i.e. each monopole carries two periodic zero-modes, as expected.
Antiperiodic zero-modes can be analyzed by looking at the $L=2$ case. 
Their distribution among
the constituent monopoles depends on the values of the $q_a$ which 
can be 0 or 1 for $m_a$ smaller or larger than $\pi$. We can distinguish three 
different cases:

$\bullet$ $\vec q = (0,0,0) $ for which $I_{\rm adj} = \sum_{a} 4 n_a$ and two periodic plus
two anti-periodic zero modes are attached to each monopole. 

$\bullet$ $\vec q = (1,0,0) $ for which $I_{\rm adj} = 8 n_1 + 2 n_2 + 2 n_3$, implying that a single 
monopole supports all 6 anti-periodic zero-modes.

$\bullet$ $\vec q = (1,1,0) $ for which $I_{\rm adj} = 6 n_1 + 6 n_2$. This is a limiting case
corresponding to one massless constituent monopole with no zero-modes, while
each of the other two, with $m=\pi$, support 4 antiperiodic zero-modes.

\noindent The previous expectation matches with our results as shown 
in Fig.~\ref{fig:su3_aper}. This example displays contour plots of the antiperiodic 
zero-mode densities for a caloron having monopoles located in the vertices of an 
equilateral triangle, with masses $m_1=m_2=m_3=2\pi/3$ (Top), and 
$m_1=4\pi/3$, $m_2=m_3=\pi/3$ (Bottom). 

\section{Closing remarks}

We want to close this manuscript  by thanking the organizers for the
invitation to contribute to this conference. Special thanks go to 
Mithat Unsal who convinced us to participate in this exciting session. 
In this talk we have  attempted to review a series of results obtained by the authors, 
putting them in perspective within a much broader  research program.
We are 
aware that the goal was too ambitious for such a short space, but we hope it 
has triggered the interest to consult the original papers and induce  
discussion.

\end{document}